\documentclass[10pt,conference]{IEEEtran} 
\IEEEoverridecommandlockouts
\usepackage{cite}
\usepackage{amsmath,amssymb,amsfonts}
\usepackage{algorithmic}
\usepackage{graphicx}
\usepackage{textcomp}
\usepackage{xcolor}
\usepackage{multirow}
\usepackage{hyperref}
\def\BibTeX{{\rm B\kern-.05em{\sc i\kern-.025em b}\kern-.08em
    T\kern-.1667em\lower.7ex\hbox{E}\kern-.125emX}}
    
\usepackage{array}
\newcolumntype{M}{>{\raggedright\arraybackslash}m{2cm}}
\newcolumntype{L}{>{\raggedright\arraybackslash}m{3cm}}

\usepackage{tikz}

\newcommand{\command}[1]{\begin{footnotesize}{\textbf{\texttt{#1}}}\end{footnotesize}}

\usepackage{ wasysym }

\begin{document}

\title{ColdPress: An Extensible Malware Analysis Platform for Threat Intelligence}

\author{
\IEEEauthorblockN{Haoxi Tan\IEEEauthorrefmark{1},
Mahin Chandramohan\IEEEauthorrefmark{2},
Cristina Cifuentes\IEEEauthorrefmark{2},
Guangdong Bai\IEEEauthorrefmark{1} and
Ryan K. L. Ko\IEEEauthorrefmark{1}}
\IEEEauthorblockA{\IEEEauthorrefmark{1}School of ITEE, The University of Queensland, Brisbane, Australia \\
Email: \{h.tan, g.bai, ryan.ko\}@uq.edu.au}
\IEEEauthorblockA{\IEEEauthorrefmark{2}Oracle Labs, Oracle, Brisbane, Australia\\
Email: \{mahin.chandramohan, cristina.cifuentes\}@oracle.com}
}

\maketitle

\begin{abstract}
Malware analysis is still largely a manual task. This slow and inefficient approach does not scale to the exponential rise in the rate of new unique malware generated. Hence, automating the process as much as possible becomes desirable. 

In this paper, we present ColdPress -- an extensible malware analysis platform  that automates the end-to-end process of malware threat intelligence gathering integrated output modules to perform report generation of arbitrary file formats.
ColdPress combines state-of-the-art tools and concepts into a modular system that aids the analyst to efficiently and effectively extract information from malware samples. 
It is designed as a user-friendly and extensible platform that can be easily extended with user-defined modules. 

We evaluated ColdPress with complex real-world malware samples~(e.g., WannaCry), demonstrating its efficiency, performance and usefulness to security analysts. Our demo video is available at \url{https://youtu.be/AwlBo1rxR1U}, and the code is open sourced on \url{https://github.com/uqcyber/ColdPress}.

\end{abstract}


\section{Introduction}
\par 
In recent years, we have witnessed the rapid rise of sophisticated cyber attacks targeting enterprises across various industry verticals.
Most of these attacks can be attributed to malware, or malicious software, which is intentionally developed to cause damage or steal information. 
According to Kaspersky, more than 24 million malware samples were reported in 2019~\cite{kaspersky}.
Despite such fast evolution of malware, malware analysis still extensively relies on manual effort to provide insight to interpret malware behaviors, and more importantly, to produce threat intelligence~(TI) for malware mitigation.  
Nevertheless, with several new malware samples released every minute, manual analysis of Indicators of Compromise (IoCs) is neither scalable nor sufficient to protect the enterprises from these large-scale malicious attacks. 

Automated and semi-automated malware analysis thus becomes highly desirable by security analysts. As such, considerable effort has been devoted to developing approaches and tools that perform malware analysis and produce IoCs based on static and dynamic code analysis \cite{vt,intelowl,hybridanalysis,Cuckoo}. 
Despite these, fundamental obstacles like tailoring the IoCs in the context of enterprise-level security still exist. 
Malware analysis has yet to be recognized as a multi-dimensional task that requires the bringing together of various views of malware to reconstruct the complete picture and context. 
Consequently, very few platforms exist to support security analysts from the perspective of \emph{integrated} intelligence. 

In this paper, we present ColdPress,  
an extensible, user-friendly and efficient pipeline that can run selected integrated analysis modules on malware samples and produce desired output formats. 
ColdPress's features include 1) a full automated solution that integrates both malware reverse engineering and TI, 2) high extensibility allowing any analysis module can be easily added without modifying the core engine, and 3) both horizontal and vertical scalability to cope with complex real-world malware samples. 
To the best of our knowledge, ColdPress is the first solution that integrates powerful Software Reverse Engineering (SRE) frameworks and TI feeds to extract information from malware.

\section{Background and Related Work}
\label{background}
In this section, we will introduce the state-of-the-art concepts and tools integrated into ColdPress as modules.

\noindent\textbf{Decompilation}. 
Decompilation aims to recover readable or even recompilable source code from a given binary. Cifuentes \cite{decompile1995} outlined the approach to do so by lifting the binary to an Intermediate Representation, analyzing it to produce a Control Flow Graph (CFG), then generating pseudocode in the target language. Today, similar approaches are used in the open source Reverse Engineering (RE)  framework Ghidra \cite{dragons} to decompile binary code into C-like source, and Radare2 \cite{r2} uses the same techniques to generate function headers and various graph outputs. The high level nature of these artifacts can help analysts understand the big picture.


\noindent\textbf{Hashing}. 
Hashing is a classic technique to identify and verify data, including files. Malware samples are often identified by their MD5 or SHA hashes. However, these hashes are checksums of the file and will change completely even when the data is a single bit off. Therefore many other useful hashing techniques had been explored before, such as fuzzy hashing \cite{ssdeep}, which does piece-wise checksums of data and therefore can detect small byte changes. 
Even more useful hashing techniques for malware detection includes hashing PE header information \cite{pehash} and hashing the CFG \cite{machoc}.

\begin{figure*}[t]
\includegraphics[width=19cm]{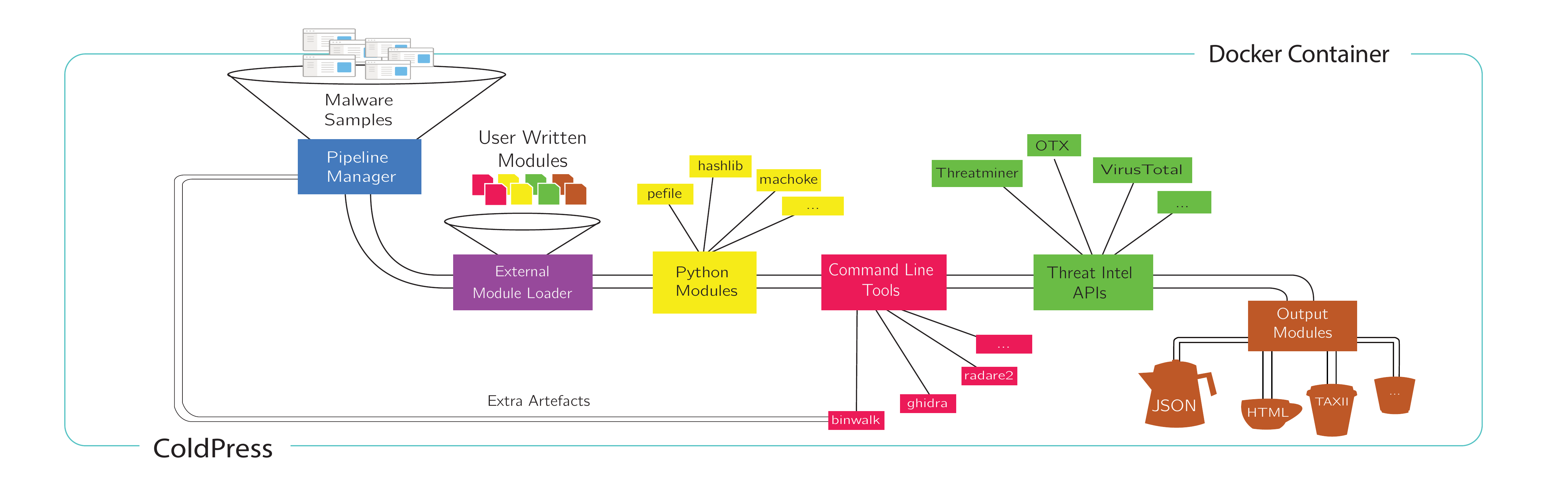}
\vspace{-0.5cm}
\caption{ColdPress Architecture}  \label{architecture}
\vspace{-0.5cm}
\end{figure*}

\noindent\textbf{Malware threat intelligence}
With the constant rise in scale of malware attacks, data sharing becomes more and more important. Malware analysis platforms such as VirusTotal \cite{vt} share TI by producing, aggregating and correlating malware IoCs for the community to use. Tools such as the MITRE ATT\&CK\cite{mitre} framework can help map information from TI feeds into actionable tactics, techniques and procedures used by criminals.

\section{ColdPress Overview} 
\subsection{Modularization}

Figure~\ref{architecture} shows the overall architecture of ColdPress. 
ColdPress is designed in a modularized manner for its extensibility. 
It consists of 1) Pipeline Manager which takes as input the malware sample(s) to analyze, and create threads for each sample, 2) External Module Loader which manages the modules programmed by the analyst, 3) Python Modules which include essential utilities such as file formatting, hashing and encoding/decoding, 4) A variety of Command Line Tools such as BinWalk \cite{binwalk} and capa \cite{capa} to run on the samples, 5) Threat Intelligence APIs queried to obtain more information such as antivirus detection and malware families, and 6) Output Modules that performs post-processing and formatting as desired by the analyst. 

The current version of ColdPress includes a set of open source libraries and tools, 
as shown in Table \ref{modules}. 
Modules marked with * are external modules - meaning that they are written as user-defined modules that do not modify the ColdPress code base, and loaded into the pipeline at run time. These also serve as templates to allow security analysts to easily add their own modules. Modules marked with \lightning \space are ``fast" modules (to be discussed in \ref{fastmode}).



\begin{table}[t]
\caption{\textbf{Integrated modules}}
\label{modules}
\resizebox{9.1cm}{!}{
    \begin{tabular}{lllll}
    Py Libs & SRE & TI APIs      & CLI tools & Output \\ \hline
    hashlib\lightning     & Ghidra\cite{dragons}         & VirusTotal\cite{vt}*\lightning  & binwalk   & JSON\lightning   \\
    machoke\cite{machoc}     & Radare2\cite{r2}\lightning        & OTX\cite{otx}*\lightning         & capa\cite{capa}      & HTML*\lightning  \\
    pefile\lightning      &                & ThreatMiner\cite{threatminer}* & YarGen*   &        \\
    pehash\cite{pehash}\lightning      &                &              &           &        \\
    ssdeep\cite{ssdeep}\lightning      &                &              &           &     \\
    regex\cite{ssdeep}\lightning      &                &              &           & 
    \end{tabular}
}
\vspace{-0.5cm}
\end{table}

\subsection {Extensibility}
\label{extensible}
ColdPress exposes an external module loader to allow users to define their own modules in Python. This allows the pipeline to be easily extended without modifying the core source code. This is inspired by the successful architectures of security testing frameworks such as BeEF \cite{beef} and Metasploit \cite{msf}. At the end of the pipeline, output modules can be added to slice a view of the output data into any file format, making it possible to automate the entire process of malware TI reporting end-to-end.

\subsection {Multi-threading}

When the amount of modules implemented into ColdPress increases, the pipeline may be delayed if those modules are ran sequentially. 
Through analyzing the analysis process, we find that only a few modules that are dependent by others need to be ran sequentially before others. 
For example, Binwalk \cite{binwalk} needs to be ran when the process starts, to extract other embedded files from a given sample before feeding them back into other modules. 
ColdPress thus runs modules other than these in parallel, to utilize the multi-core nature of modern CPUs. 

Some malware samples may take longer to run in some modules. For example, a sample with many functions and control flows would cause a path-explosion in tools such as Ghidra\cite{dragons} and capa \cite{capa}, clogging up the execution time. This is solved in ColdPress via user-defined timeouts, which can be specified per sample or in total.

All malware samples input in batch are analyzed in parallel. Theoretically, the amount of tasks in parallel $T$ would be $\mathit{T = S * M}$, 
where $S$ is number of samples and $M$ the number of loaded modules.

\subsection {Fast mode}
\label{fastmode}
ColdPress is designed to handle malware samples with batch processing. This allows a large amount of samples to be analyzed concurrently, increasing workflow efficiency.
However, the number of malware samples that can be analyzed in parallel depends on the amount of CPU power and memory available on the computer, as all modules in all samples execute in parallel by default.
To provide a lightweight analysis, ColdPress has a built-in fast mode, which runs only modules tagged as a ``fast" module. Whether a module is ``fast" or ``slow" is user-defined for optimal control of the pipeline.

\subsection{Containerization}
\label{containerization}

The system is containerized and shipped with Docker \cite{docker}. This makes ColdPress easier to deploy on different Operating Systems. This means that every time the system starts, a fresh copy of the code will be copied into a docker container with a fresh environment, and that malware files will be analyzed in an isolated environment for safety. 

By containerizing the environment \textit{and} building ColdPress to be parallel via multi-threading, the pipeline can be scaled easily both vertically (adding/removing of resources), and horizontally (by having multiple containers across multiple machines). This design is to best fit the contemporary data center and cloud-centric computing environment.

\section{Evaluation}

As ColdPress is developed to extract threat intelligence to facilitate malware analysis, our evaluation focuses on the following three research questions. 
\begin{itemize}
    \item \emph{\textbf{RQ1 (Information usefulness).} Is the intelligence ColdPress generates useful to aid the security analysts?}
    \item \emph{\textbf{RQ2 (Information quantity).} What information can ColdPress extract from the malware samples?}
    \item \emph{\textbf{RQ3 (Efficiency).} How efficiently does ColdPress work against real-world malware samples?}
\end{itemize}


\subsection {Information usefulness}


To evaluate the usefulness of ColdPress, we use it to analyze two WannaCry PE (Portable Executable) samples. 
The the identified characteristics of the malware are listed in Table \ref{wannacry_iocs}, along with the ColdPress module(s) responsible for extracting the related information. 
The evaluation shows that ColdPress can extract different types of useful information to aid understanding of the malware samples.    

\begin{table}[t]
\caption{\textbf{Identified characteristics}}
\label{wannacry_iocs}
\resizebox{9.1cm}{!}{
\begin{tabular}{l|p{1.5cm}|p{6cm}}
\hline
characteristic & CP modules              & description                                                                                                                                                                         \\ \hline
kill switch    & \vtop{\hbox{\strut regex}\hbox{\strut OTX}}               & Loose regular expression matching is used to extract strings into different categories (URL, IP, domain names and paths).   OTX also returns similar information via its various plugins.                                           \\ \hline
polymorphism   &  \vtop{\hbox{\strut hashes}\hbox{\strut pehash}\hbox{\strut machoke}} & Two different versions of the same WannaCry epoch have different MD5 but same peHash and machoke hash. 
\\ \hline
propagation    & OTX                     & OTX has a Cuckoo \cite{Cuckoo} sandbox plugin that returns dynamic analysis results when available, including network detection of WannaCry probing its neighbours \\ \hline
persistence    & capa                    & capa detects WannaCry having embedded PE files and writing to disk              \\ \hline                                                                                                    
\end{tabular}
}
\end{table}

\subsection{Information quantity}

The extracted information from ColdPress can also be used for the ``machine" use case, such as machine learning and malware clustering. Therefore, there needs to be sufficient types of data points available.

The output from ColdPress is evaluated against two other industry malware analysis platforms: IntelOwl \cite{intelowl}, a web application focused on threat intel querying, and HybridAnalysis\cite{hybridanalysis}, a dynamic malware analysis service. They are chosen because they represent mature solutions with slightly different goals than ColdPress, which integrated more software reverse engineering tools than the two. Table \ref{info_compare} compares the types of information available between ColdPress, ColdPress in fast mode, and the two platforms.

Table \ref{info_compare} clearly shows the strength of ColdPress in Reverse Engineering results compared to the other two platforms. Having a way to extract information from SRE frameworks alleviate time from manual analysis using those platforms, and although the output presentation is not as nice as inside the frameworks' UI, having extensible output formatting in ColdPress sets ground for future improvements. It is worth noting that TI API results are subject information available on those platforms, and therefore not always available in ColdPress and IntelOwl.

\begin{table}[t]
\caption{Comparison of extracted information}
\label{info_compare}
\centering
\# - Hashing, 
$\circlearrowleft$ - SRE Tools, 
$\leftrightarrows$ - Threat Intel API, 
$\circledast$ - Dynamic Analysis 


%
%
%
%
%
%

\resizebox{9cm}{!}{
    
    \begin{tabular}{p{3cm}llll}
    \hline
        & CP & CP Fast & IntelOwl   & HybridAnalysis \\ \hline
    
    \textbf{Hashes and metadata} \\ \hline
    
    md5                    & \# & \# & \# & \#     \\
    sha1                   & \# & \# &  \# &                 \\
    sha256                 & \# & \# & \#$\leftrightarrows$ & \#     \\
    ssdeep                 & \# & \# &  \# &                 \\
    pehash                 & \# & \# &            &                 \\
    machoke                & \#$\circlearrowleft$ &    &         &                 \\
    imphash                & \#$\circlearrowleft$ & \#$\circlearrowleft$ & \# &                \\
    strings                & $\circlearrowleft$ & $\circlearrowleft$ & $\circlearrowleft$ & $\circlearrowleft$$\circledast$     \\
    
    \hline \textbf{AV detection information} \\ \hline
    
    YARA generation             & $\circlearrowleft$ & &  &   $\circlearrowleft$$\circledast$              \\
    YARA detection & $\leftrightarrows$ & $\leftrightarrows$ & $\leftrightarrows$ & \\
    detected malware families           & $\leftrightarrows$  & $\leftrightarrows$ &  $\leftrightarrows$  & $\leftrightarrows$    \\
    
    MITRE ATT\&CK   & $\circlearrowleft$$\leftrightarrows$ & $\leftrightarrows$ &  $\leftrightarrows$ & $\circledast$     \\

    \hline \textbf{PE specific information} \\ \hline
    
    compilers \& packers      & $\circlearrowleft$$\leftrightarrows$ & $\circlearrowleft$$\leftrightarrows$ & $\leftrightarrows$ & $\circlearrowleft$      \\
    imports                & $\circlearrowleft$ & $\circlearrowleft$ &  $\circlearrowleft$ & $\circlearrowleft$     \\
    exports                & $\circlearrowleft$ & $\circlearrowleft$ &  $\circlearrowleft$ & $\circlearrowleft$    \\
    sections               & $\circlearrowleft$ & $\circlearrowleft$ &  $\circlearrowleft$ & $\circlearrowleft$     \\
    compile timestamps              & $\circlearrowleft$$\leftrightarrows$ & $\circlearrowleft$$\leftrightarrows$ & $\circlearrowleft$$\leftrightarrows$ & $\circlearrowleft$      \\
    
    \hline \textbf{Program semantics and functionality} \\ \hline
    
    embedded files       &  $\circlearrowleft$ & &  $\leftrightarrows$ & $\circlearrowleft$$\circledast$    \\
    symbols                & $\circlearrowleft$ & $\circlearrowleft$ &   $\leftrightarrows$         &                 \\
    function headers       & $\circlearrowleft$ & $\circlearrowleft$ &            &                 \\
    CFG     & $\circlearrowleft$ & $\circlearrowleft$ &            &                 \\
    crossref graph         & $\circlearrowleft$ & $\circlearrowleft$ &            &                 \\
    decompilation          & $\circlearrowleft$ &            &                 \\
    disassembly            & $\circlearrowleft$ &            &                 \\
    capabilities           & $\circlearrowleft$$\leftrightarrows$ & $\leftrightarrows$ & $\circlearrowleft$$\leftrightarrows$ & $\circledast$     \\

    \hline \textbf{Network related IoCs} \\ \hline
    
    IP addresses           & $\leftrightarrows$ & $\leftrightarrows$ & $\leftrightarrows$ & $\circledast$ \\
    DNS                    & $\leftrightarrows$ & $\leftrightarrows$  & $\leftrightarrows$ & $\circlearrowleft$$\circledast$     \\
    URLs                   & $\circlearrowleft$$\leftrightarrows$ & $\circlearrowleft$$\leftrightarrows$ & $\leftrightarrows$ & $\circlearrowleft$$\circledast$      \\
    
    \hline \textbf{Host related IoCs} \\ \hline
    
    windows event logs     & $\leftrightarrows$  & $\leftrightarrows$ & $\leftrightarrows$ & $\circledast$      \\
    accessed registry keys & $\circlearrowleft$$\leftrightarrows$ &  $\circlearrowleft$$\leftrightarrows$ & $\leftrightarrows$ &      $\circledast$           \\
    executed commands      & $\circlearrowleft$$\leftrightarrows$ & $\circlearrowleft$$\leftrightarrows$ & $\leftrightarrows$ &  $\circledast$      \\ \hline
    \end{tabular}
}
\end{table}

\subsection{Efficiency}

Benchmarking of ColdPress was done by putting in different amount of malware samples through the system in batch then profiling run time in seconds, CPU and memory usage.
The experiment is conducted on a Intel x64 Linux system, with the specs of Intel i5 processor with 8 logical cores and 16 GB (approx. 15GiB) of RAM. 

\begin{figure}[t]
    \includegraphics[width=7.5cm]{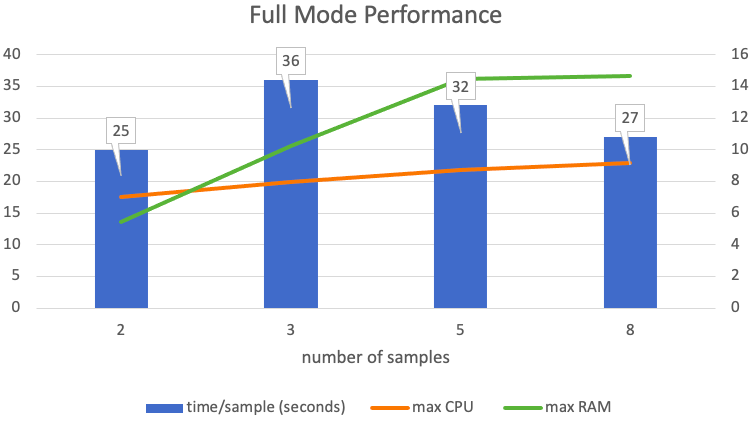}
    \includegraphics[width=7.5cm]{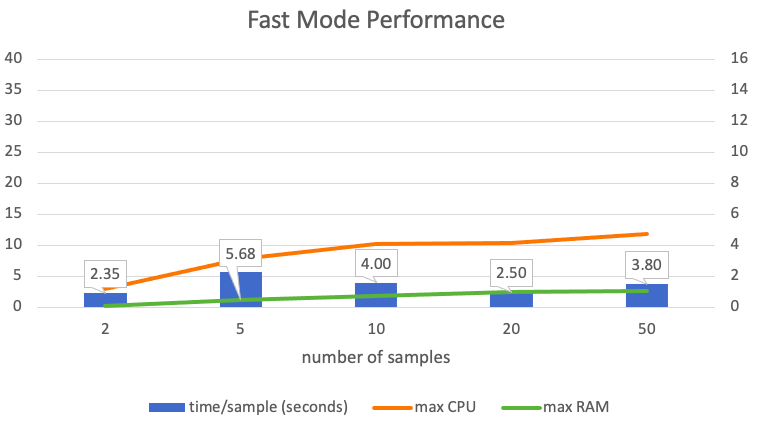}
    \caption{Performance in full and fast mode}
    \label{fig:perf}
\end{figure}

Time measurement is built into ColdPress, for while profiling of CPU and memory usage is done by running \command{docker stats} while ColdPress is running. Both full and fast mode are tested with a varying number of samples taken randomly from theZoo\cite{thezoo} dataset. The number of seconds per file, the maximum memory use in GiB and the maximum CPU use (in cores, where 1 is max utilization of a single core) are shown for both full (all modules enabled) and fast mode in Figure \ref{fig:perf}.

It is obvious to see in Figure \ref{fig:perf} that the time per sample is not affected by the number of samples analyzed in parallel (average 30s/file for full mode, and 4s/file in fast mode). However, ColdPress full mode did not scale beyond 10 samples in this experiment, as spinning up many instances of memory-intense modules such as Ghidra, which invokes the Java Virtual Machine leads to RAM being the bottleneck. ColdPress is much more scalable in fast mode, while still extracting lots of information according to Table \ref{info_compare}.

\section{ColdPress Usage}

ColdPress is written in Python and built as a Docker container. To run ColdPress, one could spawn a shell inside the Docker container \command{docker run -it coldpress bash} and then run the main script \command{run.py} inside the container.

For better usability, a shell script \command{docker\_start.sh} is available for quick spawning of the Docker container. It takes a directory as the first argument to mount into the Docker container.

To batch-analyze an entire folder of samples:
\begin{center}
    \command{./docker\_start.sh /sample/path/to/mount <args>}
\end{center}

To analyze only one sample inside a directory, assuming that the file ``filename" exists within that directory:
\begin{center}
    \command{./docker\_start.sh /sample/path/to/mount filename <args>}
\end{center}

There are many command-line switches, such as \command{-T <total timeout>}, \command{-x <m1,m2,..>} to exclude modules by name, \command{-m <m1,m2,..>} to include modules by name, and so on. They can be added at the end of the arguments. For example, to run in fast mode:

\begin{center}
    \command{./docker\_start.sh /sample/path/to/mount filename -F}
\end{center}

\section{Concluding Remarks}

In this paper, we present the design, implementation and evaluation of ColdPress, an extensible malware analysis pipeline with integrated reverse engineering tools and threat intelligence API querying. By automatically extracting numerous types of information from malware, the workflow of malware analysts has been made more efficient, and the output data can be further used for report generation and machine learning purposes.

The current version of ColdPress is limited to PE files, and it also does not perform malware de-obfuscation. In the future, we plan to extend it with more external modules, including dynamic analysis integration with sandboxes, malware unpacking APIs and more output formats.

\section*{Acknowledgments}

This project is funded by Oracle Labs through the \href{https://www.corptech.com.au/}{CEED} program.

\bibliographystyle{ieeetr}

\end{document}